\documentclass[journal=amlcef,manuscript=article]{achemso}

\usepackage[version=3]{mhchem} % Formula subscripts using \ce{}

\author{Jonathon Cottom}
\affiliation[ARCNL]
{Advanced Research Center for Nanolithography, Science Park 106, Amsterdam, 1098 XG The Netherlands}
\alsoaffiliation[UvA]
{Institute of Theoretical Physics, Institute of Physics, University of Amsterdam, Science Park 904, Amsterdam, 1098 XH The Netherlands}
\author{Qiong Cai}
\affiliation{Department of Chemical and Process Engineering, University of Surrey, Guildford, GU2 7XH United Kingdom}
\author{Emilia Olsson}
\email{k.i.e.olsson@uva.nl}
\affiliation[ARCNL]
{Advanced Research Center for Nanolithography, Science Park 106, Amsterdam, 1098 XG The Netherlands}
\alsoaffiliation[UvA]
{Institute of Theoretical Physics, Institute of Physics, University of Amsterdam, Science Park 904, Amsterdam, 1098 XH The Netherlands}

\title[An \textsf{achemso} demo]
  {Enhanced dendrite nucleation and Li-clustering at vacancies on graphene}

\begin{document}


%\begin{tocentry}
%\includegraphics{toc.jpg}

%\end{tocentry}

\begin{abstract}
An ever present challenge for Li-ion batteries is the formation of metallic dendrites on cycling that dramatically reduces cycle life and leads to the untimely failure of the cell. In this work we investigate the modes of Li-cluster formation on pristine and defective graphene. Firstly, we demonstrate that on a defect free surface the cluster formation is impeded by the thermodynamic instability of \ce{Li_2} and \ce{Li_3} clusters. In contrast, the presence of a vacancy dramatically favours clustering. This provides insights into the two modes of Li-growth observed: for the pristine basal plane if the Li-Li repulsion of the small clusters can be overcome then plating type behaviour would be predicted (rate / voltage dependent and at any point on the surface); whilst dendritic growth would be predicted to nucleate from vacancy sites, either pre-existing in the material or formed as a result of processing.
\end{abstract}

\section{Introduction}
Lithium-ion batteries (LIBs) have emerged as one of the most important electrochemical energy storage technologies in the quest for a sustainable fossil fuel-free energy future.\cite{Armand2008}  LIBs typically employ a liquid electrolyte, layered cathode and graphite anode. Despite their wide use, challenges remain in terms of  lifetime, and stability.\cite{Cai2020,Zhang2020e,Liu2020p} Fast charging LIBs are increasingly pursued which further accelerates the current LIB lifetime and stability challenges. These challenges relate to the formation of lithium dendrites at the anode surface, and irreversible lithium plating.\cite{Landi2009,Morgan2022,Liu2020a,Wu2020a} Dendrites can form and grow due to uneven deposition and accumulation of lithium at the anode electrolyte interface, leading to short-circuiting.\cite{Adams2019,Yao2022,Wang2022HighlyGrowth,Olsson2022a} For lithium plating, the lithium deposits on the anode surface at a faster rate than it can intercalate.\cite{Adams2019} This leads to reduced battery life and limited charging. As for dendrite formation, plating can also be inhomogeneous, either due to kinetics or thermodynamics, but the exact mechanism is poorly understood.\cite{WangPNAS2020} The operating voltage of graphite LIB anodes is very close to the Li metal voltage, leading to Li metal plating occurring at high charging rates or low temperatures (below 273 K).\cite{Cai2020} At extreme temperatures (above 373 K), carbon anodes are increasingly susceptible to dendrite formation.\cite{Adams2019}  

All solid state batteries (ASSBs) are widely seen as the next frontier in energy storage due to their high energy density as a consequence of the possibility of using Li metal anodes with solid electrolytes.\cite{Li2019d,Xiao2022}  With the introduction of a solid electrolyte, the expectation was that Li dendrite formation would be eradicated. However, it was found that Li dendrites are also a challenge in ASSBs.\cite{Li2019d} To this end, carbon-based materials are also finding a use in ASSBs as current collectors, protective layers, and support materials for Li metal anodes, making the atomic scale understanding of lithium nucleation on carbon surfaces imperative.\cite{Xing2021, Duan2019, Niu2019a, Yao2022,Xiao2022,LiuMitlin2019,Mukherjee2014,Gu2022Fast-chargingAerogel} 

A plethora of different carbon-materials have been employed for battery applications, including graphite, graphene, hard carbons, carbon nanotubes and soft carbons. \cite{Buiel1999,Zhao2021b,Adams2019,Li2019q} Whilst these materials are structurally different, they all contain sp$^2$ hybridized surfaces, forming the commonly present planar basal planes with connected six-membered carbon rings.\cite{Xiao2022,Mukherjee2014,Rajkamal2019,Adams2019,Olsson2022,Li2019r} In real applications, the basal plane surface can also be defective, either as a result of synthesis conditions or battery operation. Atomic scale understanding of the effect of carbon vacancy (V$_C$) on Li deposition and nucleation is crucial for the design of longer lifetime, safer battery architectures. In this letter, we conduct a density functional theory (DFT) study of Li cluster growth on graphene. In our previous work we studied the interaction of a single lithium ion with pristine and defective graphene in the context of LIBs.\cite{Olsson2019a,Olsson2021a,Olsson2021,Olsson2020,Hussain2020} These studies showed that defects have a direct impact on the single lithium binding energy and the initial lithiation during charging. Through calculations of adsorption and migration energies, we postulated that V$_C$ defects would lead to irreversible capacity loss for LIBs, a statement we here extend. Separate studies have  investigated the interaction of multiple Li atoms with different graphene models to probe the effect of lithium concentration on adsorption energy as a proxy of charging profile.\cite{Okamoto2016,Liu2014b,Liu2013b, Fan2013, Huo2019a} Small Li clusters were found to be able to adsorb more readily on graphene than Li metal (001) surface \cite{Fan2013}, and a separate study of Li clusters on pristine graphene showed that the cluster binding energy was dependent on Li concentration\cite{Liu2014b}. Here, we extend our considerations to the step wise growth of lithium clusters, exploring the energetic stability of these clusters on the pristine versus \ce{V_C} as the initial dendritic growth and plating step. 

\section{Methods}
The calculations were performed spin polarised using the CP2K\cite{Vandevondele2005,VandeVondele2007,Hutter2014,Burke1998,Kuhne2020} code with the DZVP-SR-MOLOPT\cite{VandeVondele2007} family of basis-sets to describe the valence electrons, and the GTH pseudopotentials\cite{Krack05,Goedecker96,Goedecker98} to describe the core electrons. The initial defect free relaxations were performed on a 308 atom graphene cell (see Fig S1 in the supporting information). Both the xy-lattice vectors and ion positions were fully relaxed using the quasi-Newton BFGS update scheme (ion positions only for defect calculations), the vacuum slab was converged above 10 $\textrm{\AA}$, with 25 $\textrm{\AA}$ used to account for all the tested cluster geometries.\cite{Hutter2014,Kuhne2020,Head1985,Broyden1970,Shanno1970,Fletcher1970,Goldfarb1970} The PBE functional\cite{Perdew1996A}, with the D3-BJ dispersion correction scheme\cite{Becke2005,Johnson2006,Grimme2011,Grimme2010,Grimme2007} was employed for all calculations in line with our previous work.\cite{Olsson2022a,Olsson2020,Olsson2021a,Olsson2021} The convergence criteria was 1 $\times$ 10$^{-7}$ eV, and 0.005 eV / $\textrm{\AA}$ for forces, with an energy cutoff of 750 Ry and a relative cutoff of 60 Ry to achieve an initial convergence of 0.1 meV / formula unit. To identify the lowest energy cluster configurations for each size, an exhaustive sampling was performed starting from the single atom (\ce{Li_1}). In each case the identified lowest energy configuration becomes the starting point for \ce{Li_{(n+1)}}. The process is iterated for the cluster n = 1 - 12 presented here. For completeness, the results for the higher energy cluster configurations are  included in the supplementary information Fig S2-3. The interaction energies are calculated using the standard formalism of Zhang and Northrup\cite{Zhang1991a} in terms of formation energy \ce{E_f} (eq. \ref{eq:Ef}), binding energy (\ce{E_{bind}}) (eq. \eqref{eq:Ebind}), and cohesive energy (\ce{E_{coh}}) (eq. \eqref{eq:Ecoh}).

\begin{equation} \label{eq:Ef}
   E_f = \frac{E_{Li_n@surface} - E_{surface} - {n \mu_{Li}}}{n} 
\end{equation}

\begin{equation}\label{eq:Ebind}
   E_{bind} = \frac{E_{Li_n@surface} - E_{surface} - E_{Li_n}}{n}
\end{equation}

\begin{equation}\label{eq:Ecoh}
   E_{coh} = \frac{E_{Li_n} - {n \mu_{Li}}}{n}
\end{equation}

Here, ${E_{Li_{n@surface}}}$ is the total energy of the $Li_n$ cluster on the surface, $E_{surface}$ is the total energy of the surface, n the number of Li, and $\mu_{Li}$ the lithium chemical potential. The Li chemical potential was varied from the vacuum ($\mu_{Li}$ defined as a single Li atom in vacuum) to metallic reference ($\mu_{Li}$ defined from Li bulk).  

\section{Results and Discussion}
\begin{figure}[htp]
    \centering
    \includegraphics[width=0.9\textwidth]{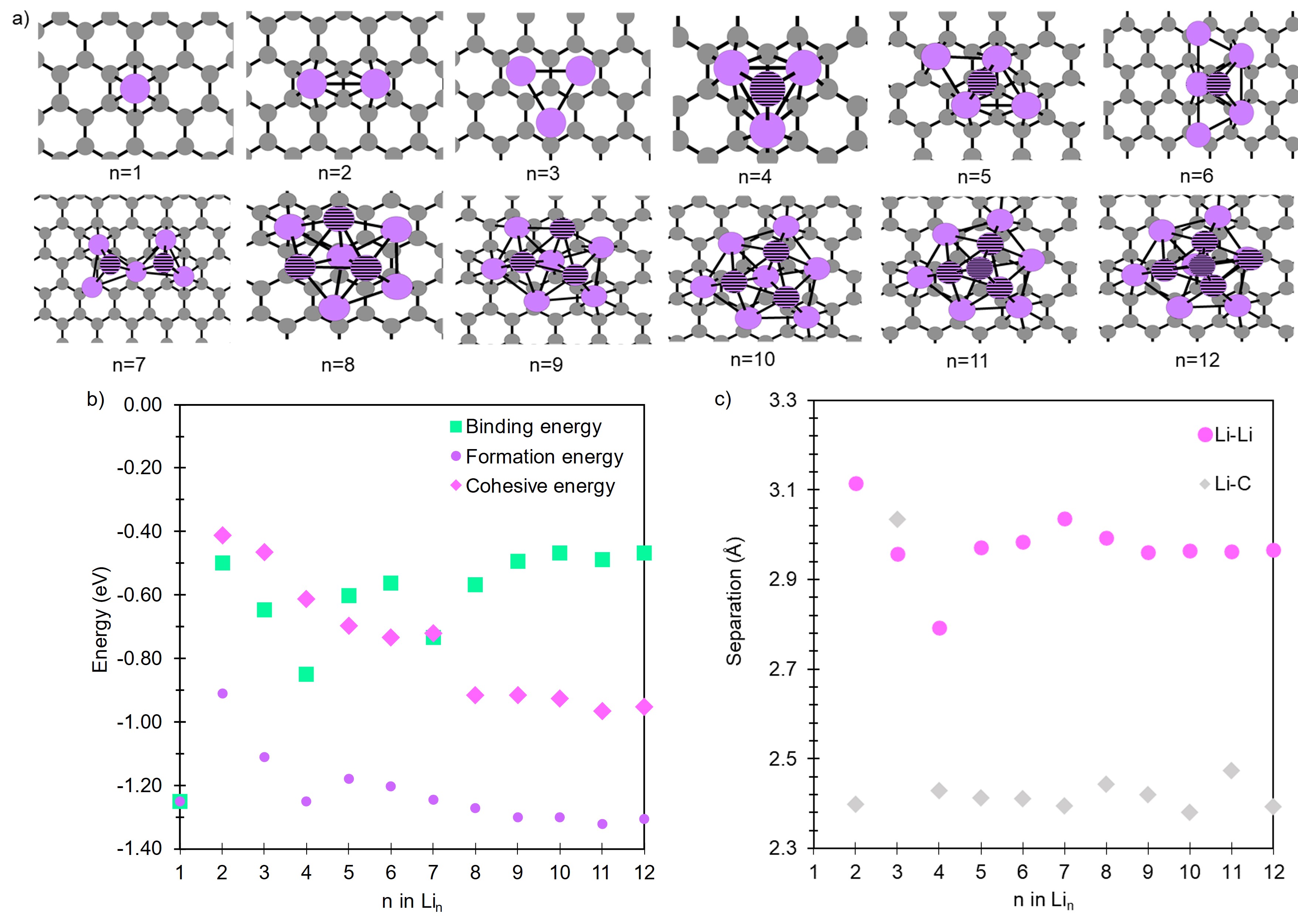}
    \caption{a) lowest energy cluster configurations for the range n = 1 - 12 on the pristine where purple circles are Li (striped purple circles denotes Li in next layer), and grey C. Figures made using the Atomic Simulation Environment (ASE).\cite{HjorthLarsen2017} b) the interaction energies decomposed into (\ce{E_{bind}}), (\ce{E_{coh}}) and (\ce{E_{f}}), and Li-C and Li-Li separation (c). }
    \label{Fig1}
\end{figure}

To explore how \ce{V_C} impact the Li nucleation on graphene, two key questions need to be addressed. Firstly, how the presence of a vacancy impacts \ce{E_{bind}} of successive Li atoms with respect to the isolated Li, in essence how is surface-Li binding aided or impeded by the presence of a \ce{V_C} defect? Secondly, how does the presence of the defect influence the morphology and \ce{E_{coh}} of the forming clusters? Probing whether the cluster formation is significantly aided or frustrated on forming at a vacancy site. To act as a reference, cluster binding is first calculated on the defect free basal plane, Li-atoms are added step-wise to the lowest energy sites identified in ref\cite{Olsson2019a}, \ce{E_{bind}}, \ce{E_{coh}} and \ce{E_{f}}  are then extracted (Fig.\ref{Fig1}). 

It is clear from the configurations in Figure \ref{Fig1}a, that the preferred hole site absorption geometry of the isolated Li, results in a significant distortion for clusters from n $\geq$ 2. Each Li cannot be accommodated at the high symmetry hole site, as a result it is distorted from its central position. For the smallest clusters (n = 2 and 3) the repulsive energetic penalty can be estimated for the induction of a single neighbouring Li as 0.34 eV. This is expressed in both \ce{E_{bind}} (Figure \ref{Fig1}b) and the C-Li separation with the cluster moving away from the surface (Figure \ref{Fig1}c). The repulsive interaction is partially compensated for by the Li-Li \ce{E_{coh}} (Figure \ref{Fig1}b), which shows a marked increase with increasing cluster size. For clusters n $>$ 3 the addition of the (n+1)$^{th}$ Li to the next layer becomes favoured. The cluster then grows stepwise via the building up of node sharing trigonal pyramidal units. The ground state cluster geometries and relative energies are in agreement with the results from Liu and co-workers.\cite{Liu2014b}

For the Li clusters explored, \ce{E_{bind}} rapidly decreases and converges to -0.45 eV / atom, at the same rate \ce{E_{coh}} of the cluster increases to -0.91 eV. As the cluster structure forms layers, a further differentiation between those atoms that interact with the C-surface and those that only interact with Li is seen. This is expressed in the partial charges for the larger clusters with a clear negative shift of the Mulliken charge for the non-surface interacting Li atoms (+0.02 e) when compared to the surface interacting Li in a given cluster (+0.47 e) and the isolated case (+0.5 e). It is important to note that for the defect free system the small clusters (n $<$ 3) are energetically unfavoured with respect to the dispersed Li, additionally clusters in the n =4 - 6 are iso-energetic with the dispersed Li. Therefore, clustering would only be predicted to be stable for the larger clusters (n $\geq$ 7).

\begin{figure}[ht]
    \centering
    \includegraphics[width=0.9\textwidth]{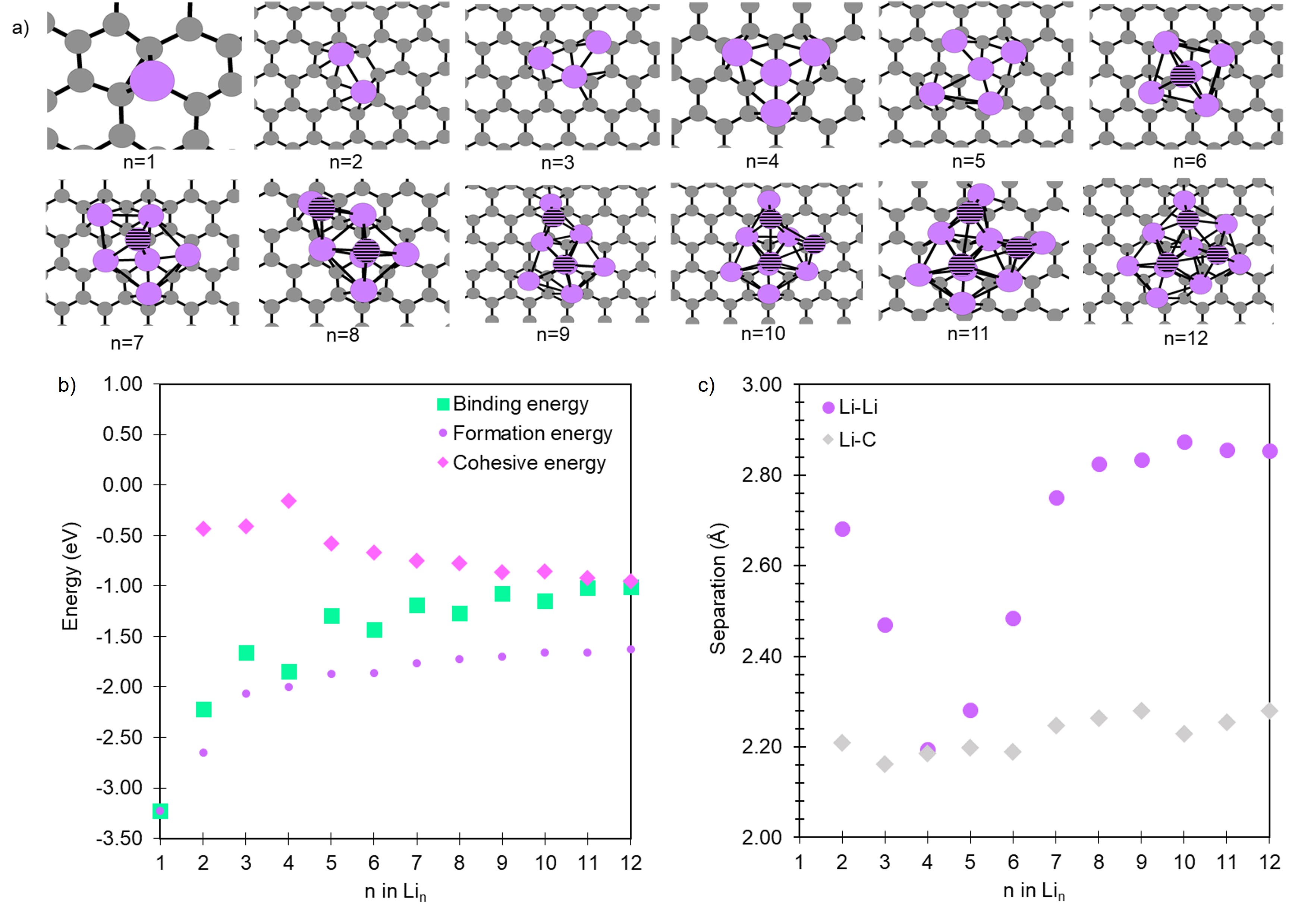}
    \caption{a) lowest energy cluster configurations for the range n = 1 - 12 at the \ce{V_C} where purple circles are Li (striped purple spheres denotes Li in next layer), and grey C. Figures were made using ASE.\cite{HjorthLarsen2017} b) the interaction energies decomposed into (\ce{E_{bind}}), (\ce{E_{coh}}) and (\ce{E_{f}}). c) The mean cluster Li-C and Li-Li separations (c).}
    \label{Fig2}
\end{figure}

Figure \ref{Fig2}a shows the lowest energy of Li configurations for binding on the \ce{V_C}. In agreement with the previously reported isolated Li\cite{Olsson2019a}, \ce{E_{bind}} is -3.3 eV and the Li interacts with the under-coordinated C-atom, while the other 2 vacancy C-atoms form a 5-membered ring. This geometry remains stable for n = 1 \& 2, however for clusters n $\geq$ 3 the 5-membered ring opens, partially for n=3 and completely at n = 4 resulting in a Li sitting at the central C$_3$ site (Figure \ref{Fig2}a). The under-coordinated C-atoms show a significant distortion lifting 1.2 $\textrm{\AA}$ out of the plane. This can clearly be seen in the Li-Li separation (Figure \ref{Fig2}c), with a decrease from n=2 to n=4, followed by a sharp increase once the defect and adjacent sites are occupied. Above n = 4 the next Li can either be added adjacent to the cluster or on top forming a new Li-layer. Similar to metal adsorption on the basal plane, the metal top site is energetically favoured. The cluster then grows in lateral extent by the addition of the next Li to the adjacent hole site in a manner similar to that seen on the basal plane. The presence of a vacancy leads to a marked increase in \ce{E_{f}} and \ce{E_{bind}} for the smallest clusters (n = 2 - 5). \ce{E_{coh}} as a function of cluster size is of the same order for both the \ce{V_C} and pristine, being a function of cluster size and the effective coordination that this supports. As the cluster increases in size \ce{E_{bind}} will converge to that of the pristine (the influence of the vacancy bound Li becomes negligible in per atom terms). Whereas \ce{E_{coh}} will converge towards Li-metal (-1.78 eV / atom) for clusters of sufficient size. The presence of the vacancy in essence anchors the cluster in place, allowing the repulsive interactions seen in the small clusters on the pristine to be overcome. Hence cluster nucleation would be predicted to be favourable at the \ce{V_C} for all of the cluster sizes considered.

It is clear that in both the pristine and the \ce{V_C} there is a propensity to form clusters by preferentially adding Li-atoms to the next layer for clusters greater than n=3 and n=4, respectively. In each case it is most energetically favourable to grow the cluster \textit{via} filling the adjacent hole sites, followed by the addition to the top site. This leads to the stepwise growth of the cluster, for the range of cluster sizes considered here the growth is limited to 2- / 3-layers. The main difference between the two binding modes is the magnitude of the surface binding energy, with binding at the vacancy dramatically favoured. It should be noted that the reduction in the cluster binding energy for the pristine is largely driven by strain between neighbouring Li with the system attempting to minimise repulsive Li-Li interaction, while maximising Li-surface interaction. Here the average picture given by \ce{E_{bind}} gives a reasonable picture of cluster binding as each site is approximately equivalent. In the \ce{V_C} system the situation is complicated as there is a large discrepancy in \ce{E_{bind}} between sites\cite{Olsson2019a}, in this case the vacancy site (n = 1), and the filling of the adjacent sites (n = 2 - 4) are in essence anchored in position interacting strongly with the vacancy site(s). As the cluster grows the available sites for the next Li begin to resemble the pristine system and this is seen in the increase in C-Li separation (Figure \ref{Fig2}c), which approaches the pristine values as the number of pristine sites dominate.  

\begin{figure}[h]
    \centering
    \includegraphics[width=0.5\textwidth]{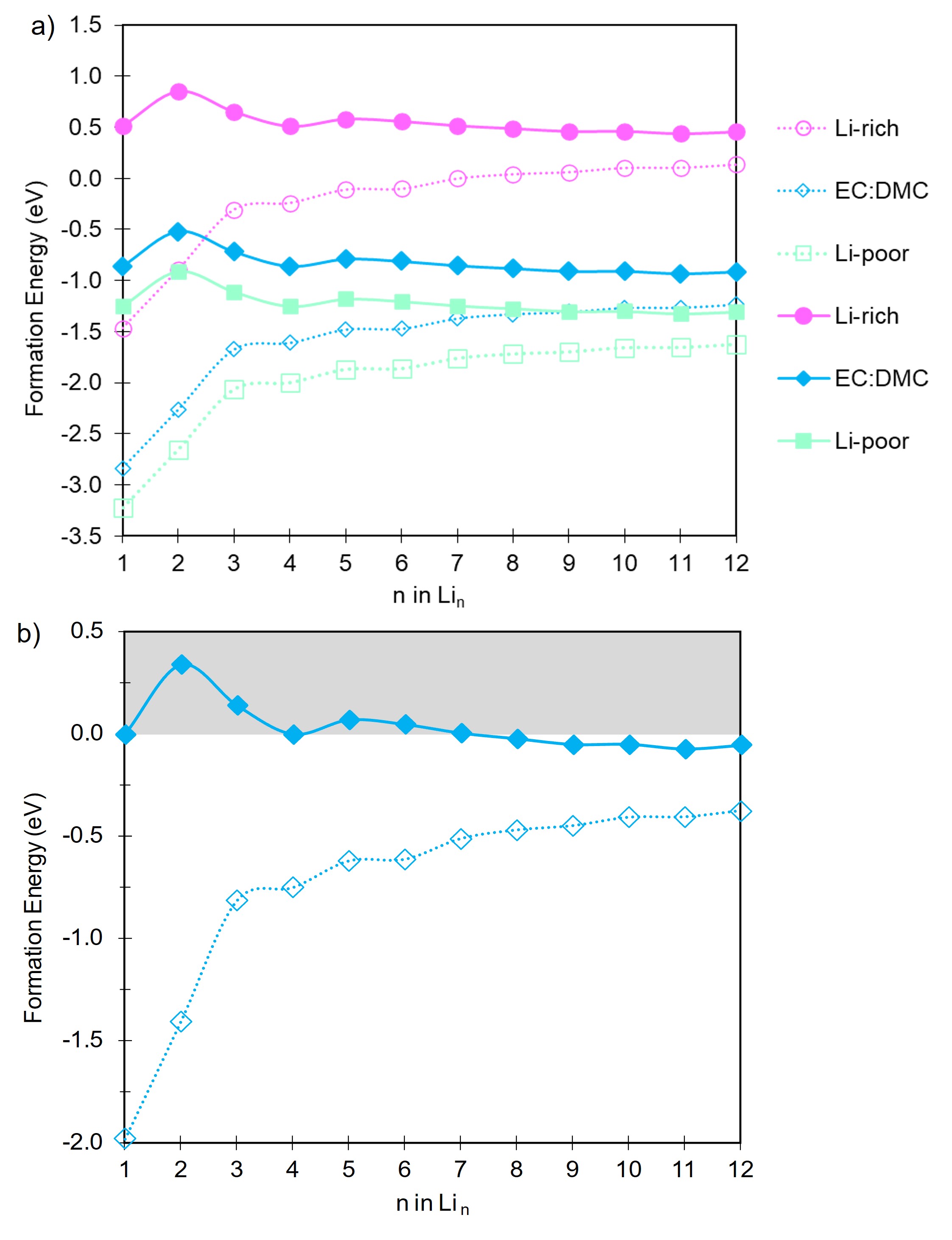}
    \caption{a) Li cluster formation energy on pristine (filled markers and non-dotted lines) and defective (empty markers and dotted lines) pristine with the Li energy taken at the Li-poor, Li-rich, and Li solvated in EC:DMC ([3:1]) chemical potential. b) shows the lithium cluster formation energy with the solvated Li chemical potential, referenced to a single Li bonded to the pristine surface from the solvated Li reference.}
    \label{Fig3}
\end{figure}

Figure \ref{Fig3}a shows \ce{E_{f}}  of the Li clusters on pristine and defective pristine (\ce{E_{f}} in Figure \ref{Fig1}b and \ref{Fig2}b) as a function of Li chemical potential. For Li-poor conditions, the chemical potential is taken from the vacuum reference, Li-rich from the Li metal reference, and solvated Li is Li in 3:1 EC:DMC as described by Fan et al. (2019).\cite{Fan2019NatEn}. Taking the energy of a single Li on the pristine in the solvated Li assumption as reference, it can be seen from Figure \ref{Fig3}b that only the larger Li clusters are energetically stable on the pristine basal plane, and formation of the small clusters (n $<$ 4) is thermodynamically hindered. Upon the introduction of a \ce{V_C}, all clusters become thermodynamically favoured. 

The pristine and \ce{V_C} show a number of similarities and some important differences upon the formation of Li clusters. In both cases stepwise growth is observed, with  adjacent hole sites (\ce{Li_2} and \ce{Li_3}) being filled before adding a new layer (\ce{Li_4}). Cluster growth on the pristine is significantly hampered for the smallest cluster n=2-4, which are 0.25 eV/atom less favoured than the dispersed Li. This suggests that cluster growth will only be observed at high Li-concentrations / -rates, and is potentially linked to Li-plating behaviour reported. In contrast the presence of the \ce{V_C} dramatically increases the binding strength for the small clusters, the formation of which is unfavourable on the pristine surface. In addition, this strong binding anchors the growing cluster onto the vacancy site. As the cluster size increases the available site resemble the pristine basal plane, reducing the effect of the \ce{V_C}.  

\section{Conclusion}

In summary, cluster growth is explored on both the pristine basal plane and the \ce{V_C} system. Cluster formation in the pristine system is hampered by the thermodynamic instability of the small Li-clusters with respect to disperse Li adsorption. Therefore, clustering would only be observed at high deposition rates (Li-concentration). This mode of growth would be homogeneous across the surface and analogous to the Li-plating behaviour observed. In contrast the presence of the \ce{V_C} results in a dramatic increase in Li-binding energy, facilitated by the relaxation of the cluster to strongly bind the n = 4 cluster. The result of this is two-fold firstly, the thermodynamic instability of the small clusters seen on the pristine surface is overcome. Secondly, the forming cluster is anchored in place at the vacancy site. This results in inhomogeneous growth mediated by surface vacancies resulting in dendritic growth. Further investigations are underway to characterise those additional defect types that favour, or indeed hamper clustering.

\begin{acknowledgement}

 This work made use of the Dutch national e-infrastructure with the support of the SURF Cooperative using grant no. EINF-2434. The authors thank SURF (www.surf.nl) for the support in using the Lisa Compute Cluster and National Supercomputer Snellius. Q.C. would like to acknowledge funding support from Faraday Institution through the LiSTAR programme (EP/S003053/1, Grant FIRG014), and funding support from Horizon Europe through the OPERA consortium (Grant No. 101103834). E.O. is grateful for a WISE Fellowship from the Dutch Research Council (NWO). This work has been carried out at the Advanced Research Center for Nanolithography (ARCNL). ARCNL is a public–private partnership with founding partners UvA, VU, NWO-I, and ASML and associate partner RUG. 
\end{acknowledgement}

\begin{suppinfo}

Higher energy Li cluster configurations are included in the supplementary information.
\end{suppinfo}

\bibliography{references}

\end{document}